\documentclass[sigconf]{acmart}

\newcommand{\ToolName}{\sc EthCRAFT}

\newcommand{\NumClient}{five}
\usepackage{amsmath,amsfonts}
\usepackage{algorithmic}
\usepackage{graphicx}
\usepackage{textcomp}
\usepackage{makecell}
\usepackage{multirow}
\usepackage{enumitem}
\usepackage[linesnumbered,ruled]{algorithm2e}
\usepackage{listings}
\usepackage{pifont}
\usepackage{colortbl}
\definecolor{mygray}{gray}{0.95}
\definecolor{magicmint}{rgb}{0.85, 0.96, 0.92}
\usepackage{tikz}
\newcolumntype{P}[1]{>{\centering\arraybackslash}p{#1}}

\definecolor{cgreen}{RGB}{106,153,85}
\definecolor{corange}{RGB}{245,198,60}

\newcommand\SlashCell[2]{%
  \begin{tikzpicture}[baseline=(current bounding box.center)]
    \node[minimum width=2cm, minimum height=0.8cm] (box) {};
    \draw (box.south east) -- (box.north west);
    \node[anchor=south west, font=\bfseries] at (box.south west) {#1};
    \node[anchor=north east, font=\bfseries] at (box.north east) {#2};
  \end{tikzpicture}%
}

\definecolor{NGreen}{RGB}{80,176,80}
\definecolor{verylightgray}{rgb}{.97,.97,.97}
\lstdefinelanguage{Solidity}{
  keywords=[1]{anonymous, assembly, assert, balance, break, call, callcode, case, catch, class, constant, continue, constructor, contract, debugger, default, delegatecall, delete, do, else, emit, event, experimental, export, external, false, finally, for, function, gas, if, implements, import, in, indexed, instanceof, interface, internal, is, length, library, log0, log1, log2, log3, log4, memory, modifier, new, payable, pragma, private, protected, public, pure, push, require, return, returns, revert, selfdestruct, send, solidity, storage, struct, suicide, super, switch, then, this, throw, true, try, typeof, using, value, view, while, with, addmod, ecrecover, keccak256, mulmod, ripemd160, sha256, sha3}, 
  keywordstyle=[1]\color{blue}\bfseries,
  keywords=[2]{address, bool, byte, bytes, bytes1, bytes2, bytes3, bytes4, bytes5, bytes6, bytes7, bytes8, bytes9, bytes10, bytes11, bytes12, bytes13, bytes14, bytes15, bytes16, bytes17, bytes18, bytes19, bytes20, bytes21, bytes22, bytes23, bytes24, bytes25, bytes26, bytes27, bytes28, bytes29, bytes30, bytes31, bytes32, enum, int, int8, int16, int24, int32, int40, int48, int56, int64, int72, int80, int88, int96, int104, int112, int120, int128, int136, int144, int152, int160, int168, int176, int184, int192, int200, int208, int216, int224, int232, int240, int248, int256, mapping, string, uint, uint8, uint16, uint24, uint32, uint40, uint48, uint56, uint64, uint72, uint80, uint88, uint96, uint104, uint112, uint120, uint128, uint136, uint144, uint152, uint160, uint168, uint176, uint184, uint192, uint200, uint208, uint216, uint224, uint232, uint240, uint248, uint256, var, void, ether, finney, szabo, wei, days, hours, minutes, seconds, weeks, years},  
  keywordstyle=[2]\color{teal}\bfseries,
  keywords=[3]{block, blockhash, coinbase, difficulty, gaslimit, number, timestamp, msg, data, gas, sender, sig, value, now, tx, gasprice, origin},  
  keywordstyle=[3]\color{violet}\bfseries,
  identifierstyle=\color{black},
  sensitive=false,
  comment=[l]{//},
  morecomment=[s]{/*}{*/},
  commentstyle=\color{gray}\ttfamily,
  stringstyle=\color{red}\ttfamily,
  morestring=[b]',
  morestring=[b]"
}
\usepackage{tikz}
\newcommand*\circled[1]{\tikz[baseline=(char.base)]{
		\node[shape=circle,draw,inner sep=1pt, fill=black, text=white] (char) {#1};}}

\ifodd 0
\newcommand{\zj}[1]{\textcolor{cgreen}{{#1}}} 
\else
\newcommand{\zj}[1]{#1}
\fi

\ifodd 0
\newcommand{\zjrv}[1]{\textcolor{cgreen}{{#1}}} 
\else
\newcommand{\zjrv}[1]{#1}
\fi

\AtBeginDocument{%
  }

\begin{document}

\title{Is My RPC Response Reliable? Detecting RPC Bugs in Ethereum Blockchain Client under Context}


\author{Zhijie Zhong}
\affiliation{%
  \department{School of Software Engineering, GuangDong Engineering Technology Research Center of Blockchain}
  \institution{Sun Yat-sen University}
  \city{Zhuhai}
  \country{China}
}
\orcid{0000-0002-2427-0641}
\email{zhongzhj3@mail2.sysu.edu.cn}

\author{Yuhong Nan}
\affiliation{%
  \department{School of Software Engineering, GuangDong Engineering Technology Research Center of Blockchain}
  \institution{Sun Yat-sen University}
  \city{Zhuhai}
  \country{China}
}
\orcid{0000-0001-9597-9888}
\email{nanyh@mail.sysu.edu.cn}

\author{Mingxi Ye}
\authornote{corresponding author}
\affiliation{%
  \department{School of Software Engineering, GuangDong Engineering Technology Research Center of Blockchain}
  \institution{Sun Yat-sen University}
  \city{Zhuhai}
  \country{China}
}
\orcid{0009-0004-6708-4074}
\email{yemx6@mail2.sysu.edu.cn}

\author{Qing Xue}
\affiliation{%
  \institution{Sun Yat-sen University}
  \city{Guangzhou}
  \country{China}
}
\orcid{0009-0002-5625-0036}
\email{xueq25@mail2.sysu.edu.cn}

\author{Jiashui Wang}
\affiliation{%
  \institution{Zhejiang University}
  \city{Hangzhou}
  \country{China}
}
\orcid{0009-0005-3100-0534}
\email{12221251@zju.edu.cn}

\author{Long Liu}
\affiliation{%
  \institution{Independent Researcher}
  \city{Hangzhou}
  \country{China}
}
\orcid{0009-0000-5032-8475}
\email{lvbluesky@qq.com}

\author{Xinlei Ying}
\affiliation{%
  \institution{Independent Researcher}
  \city{Hangzhou}
  \country{China}
}
\orcid{0009-0007-2082-863X}
\email{0x140ce@gmail.com}

\author{Zibin Zheng}
\affiliation{%
  \department{School of Software Engineering, GuangDong Engineering Technology Research Center of Blockchain}
  \institution{Sun Yat-sen University}
  \city{Zhuhai}
  \country{China}
}
\orcid{0000-0002-7878-4330}
\email{zhzibin@mail.sysu.edu.cn}

\begin{abstract}

  Blockchain clients are fundamental software for running blockchain nodes. They provide users with various RPC (Remote Procedure Call) interfaces to interact with the blockchain. These RPC methods are expected to follow the same specification across different blockchain nodes, providing users with seamless interaction. However, there have been continuous reports on various RPC bugs that can cause unexpected responses or even Denial of Service weaknesses. Existing studies on blockchain RPC bug detection mainly focus on generating the RPC method calls for testing blockchain clients. However, a wide range of the reported RPC bugs are triggered in various blockchain contexts. To the best of our knowledge, little attention is paid to generating proper contexts that can trigger these context-dependent RPC bugs.
  
  In this work, we propose {\ToolName}, a \underline{C}ontext-aware \underline{R}PC \underline{A}nalysis and \underline{F}uzzing \underline{T}ool for client RPC bug detection. {\ToolName} first proposes to explore the state transition program space of blockchain clients and generate various transactions to construct the context. {\ToolName} then designs a context-aware RPC method call generation method to send RPC calls to the blockchain clients. The responses of {\NumClient} different client implementations are used as cross-referencing oracles to detect the RPC bugs. We evaluate {\ToolName} on real-world RPC bugs collected from the GitHub issues of Ethereum client implementations. Experiment results show that {\ToolName} outperforms existing client RPC detectors by detecting more RPC bugs. Moreover, {\ToolName} has found six new bugs in major Ethereum clients and reported them to the developers. One of the bug fixes has been written into \textit{breaking changes} in the client's updates. \zjrv{Three of our bug reports have been offered a vulnerability bounty by the Ethereum Foundation.}


\end{abstract}



\maketitle

\section{Introduction}
\label{sec:intro}

Blockchain has established an enormous software ecosystem in the past few years. As decentralized ledgers, the blockchain runs on a network of blockchain nodes, each of which maintains an individual copy of the blockchain states. These nodes run on specialized software programs, specifically blockchain execution clients (referred to as blockchain clients in this paper). Currently, there are several blockchain client implementations for the Ethereum blockchain network, one of the most popular blockchains. These clients are implemented in various program languages, including Go~\cite{geth,erigon}, Java~\cite{besu}, and Rust~\cite{reth}. They serve as the critical infrastructure of blockchain systems, as the whole blockchain network runs on them. Once a bug is found in a blockchain client, all blockchain nodes using this client implementation will be affected.

As the interface between blockchain users and blockchain systems, blockchain clients provide a series of standardized Remote Procedure Call (RPC) methods for interacting with blockchains. These RPC methods offer a wide range of functionalities and are used in various situations. For example, blockchain wallets use these RPC methods to send users' transactions to the blockchain. Besides, a large number of decentralized applications are reported to rely on the RPC interfaces to support their off-chain front-end functions~\cite{kim2023etherdiffer}. While the blockchain client implementation is generally reliable, the reliability of RPC methods remains a problem of concern. Many issues regarding the RPC bugs have been raised within the major blockchain clients. These bugs can cause the RPC to return incorrect blockchain states and even result in a Denial of Service (DoS) attack against the clients.

Motivated by these problems, Ethereum Foundation, the official Ethereum group, has launched several developer agendas to improve the standardization and reliability of Ethereum client RPC services~\cite{RPCstandard}. Several studies~\cite{yi2022empirical} have also been proposed to study RPC bugs in blockchain clients. Li et al.~\cite{li2021strong} and Luo et al.~\cite{luo2024towards} studied exploit construction methods for a specific type of RPC DoS bug based on manually defined rules, thereby lacking scalability. EtherDiffer~\cite{kim2023etherdiffer} proposed a fuzzing framework for RPC bug detection, but they only detect RPC bugs in static environments, i.e., with fixed blockchain states and limited transactions that are manually constructed. However, there has been a rising number of RPC bug reports where the bugs can only be triggered under specific blockchain contexts, e.g., when certain transactions or account states are present. Such RPC bugs can undermine the usability and reliability of blockchain clients. Yet, existing works fall short of automatically detecting such context-dependent RPC bugs, as they cannot generate flexible contexts to trigger them. 

For example, a widely used Ethereum client, Hyperledger Besu~\cite{besu}, was reported to have a Denial of Service weakness~\cite{ethcallissue} in its transaction simulation RPC method, including \texttt{\small eth\_call, trace\_call} and \texttt{\small trace\_callMany}. These RPC methods allow users to simulate the execution of their transactions on a blockchain client without sending these transactions on-chain. Due to a defective configuration, Hyperledger Besu allows users to utilize unlimited computational resources of a node to simulate transactions by default. Consequently, adversaries can send these RPC requests with specific parameters to cause DoS attacks on the blockchain node. Notably, this bug can only be triggered under two conditions: the blockchain context must contain a specific smart contract, and the adversary must set a specific parameter for the RPC method call.

\textbf{Challenges.} Detecting such bugs is non-trivial due to the following distinct challenges: \textbf{1) Context Generation.} The context-dependent RPC bugs are triggered under specific blockchain states and transactions, i.e., blockchain contexts. The blockchain context space is extremely huge as it contains an exponential combination of account states, contract operations, and transactions. Therefore, it is challenging to generate the context of interest that can trigger RPC bugs. \textbf{2) Method Call Generation.} RPC bug detectors need to establish efficiency in exploring the RPC method call space. Blockchain clients support more than 40 RPC methods, and each of them allows a wide range of parameter inputs, including pre-constructed transactions, dynamic arrays, etc. Therefore, it is also challenging to generate fine-grained parameters for these RPC methods to trigger bugs. \textbf{3) On-chain Execution Overhead.} To ensure a consistent blockchain context for testing diverse client implementations, a common choice is to deploy these clients on a test blockchain network. However, this approach also incurs considerable computational overhead and delays regarding the auxiliary blockchain modules, such as consensus and signature verification. To trigger the context-dependent RPC bugs, various transactions need to be deployed on the blockchain during the context generation and mutation process. Thus, the computational overhead is further exaggerated in context-dependent RPC bug detection.


\textbf{Our work.} In this work, we propose {\ToolName}\footnote{availiable at \url{https://github.com/Z-Zhijie/EthCRAFT}.}, a \underline{C}ontext-aware \underline{R}PC \underline{A}nalysis and \underline{F}uzzing \underline{T}ool for client RPC bug detection. To address the aforementioned challenges, {\ToolName} features a decoupled design with two key modules: off-chain context space exploration and on-chain method call generation. In the off-chain context space exploration module, {\ToolName} generates various blockchain contexts via transaction mutation. We design an efficient transaction selection method to select from millions of Ethereum mainnet transactions as the initial corpus. Runtime state-aware mutation strategies are proposed to mutate the opcode sequences of executed transactions. To support the off-chain transaction evaluation, {\ToolName} designs a transaction execution simulator based on the Go-Ethereum client, which allows {\ToolName} to directly execute and evaluate a transaction without invoking irrelevant modules. In the on-chain method call generation module, {\ToolName } establishes a test blockchain network with transactions of interest as context. The test blockchain network consists of {\NumClient} diverse Ethereum client implementations. We propose a context-aware method for generating RPC calls. RPC responses across different clients are used as cross-referencing oracles for bug detection. 

We evaluate the proposed {\ToolName} on a real-world dataset collected from the GitHub issues of {\NumClient} Ethereum client implementations. The evaluation result shows that {\ToolName} outperforms existing RPC bug detectors by detecting more bugs on the dataset. Moreover, we report six new RPC bugs in the major Ethereum clients, all of which were confirmed by the developers.

The contributions of the paper are outlined as follows:
\begin{itemize}[leftmargin=*]
    \item We propose a novel approach for detecting the context-dependent RPC bugs for Ethereum blockchain clients. We design a set of novel mechanisms to improve the efficacy and efficiency of RPC bug detection.
    \item We propose an effective method for generating blockchain contexts. By leveraging initial transaction corpus selection and runtime state-aware mutation, we can efficiently explore the transaction execution program space of the blockchain client. 
    \item We propose a decoupled framework for the context generation and RPC testing process. Based on this framework, we can improve the efficiency of the proposed detector. 
    \item We evaluate {\ToolName} on a real-world dataset consisting of 30 reported RPC bugs. {\ToolName} outperforms prior works by detecting more bugs. We find and report six new bugs to the developers. One of them is written into \textit{breaking changes} in a regular update of a major Ethereum client~\cite{BesuBreakingchanges}. Three of them have been offered a vulnerability bounty from Ethereum Foundation.
\end{itemize}

\section{Background}
\label{sec:background}

\subsection{Ethereum Blockchain Clients.}
Ethereum is one of the most popular blockchain platforms. It is the first blockchain to provide a blockchain virtual machine, i.e., Ethereum Virtual Machine (EVM), which supports deploying and executing programs on the blockchain. These programs running on the blockchain are called smart contracts. With the help of smart contracts, developers can deploy diverse applications with complex business logic on top of the blockchain.

The Ethereum network is composed of decentralized blockchain nodes, each running two types of clients: the consensus client and the execution client. To prevent the network from halting due to bugs in a particular client, Ethereum has been working on increasing the diversity of client implementations, which are developed and maintained in different program languages. These clients are developed based on the Ethereum specifications and implement the same functionality. 


This work focuses on the execution client. 
The client is mainly composed of four modules: storage database, transaction processing, networking, and interface processor. 
Each execution client maintains an individual copy of the blockchain states. Once a new block is added to the blockchain, the client first executes the transactions in the block based on the transaction processing module. For each transaction, the client retrieves the related account states from the storage database and invokes the Ethereum Virtual Machine (EVM) implementation to execute the transaction's instructions. The client also maintains its own transaction pool for listening to the transactions to be packed into new blocks. Beyond these modules, an execution client provides an interface layer that includes the RPC interface and engine API interface. The lower-level engine API interface interacts with blockchain consensus clients. The RPC interface implements processors for each RPC request from blockchain users and applications.


\subsection{RPC methods in Blockchain Client.}
Ethereum clients provide various RPC methods. The input and expected behavior of these methods are specified in the Ethereum official RPC documents~\cite{RPCspec} and the client documents~\cite{Clientspec}.

Overall, these RPC methods mainly provide four types of functionalities: 1) Transaction sending. These RPC methods can change the global blockchain states by sending on-chain transactions. For example, \texttt{\small eth\_sendTransaction} takes a signed transaction as input and submits the transaction to the transaction pool of the blockchain client and its peers; 2) Blockchain information retrieval. These RPC methods allow users to request the global information of the blockchain. For example, the \texttt{\small eth\_getBalance} method returns the balance of a given address; 3) On-chain transaction decoding. These RPC methods can decode a given transaction on the blockchain and return the executed EVM operations and the EVM state during transaction execution. For example, \texttt{\small debug\_traceTransaction} takes in an on-chain transaction identifier and outputs the trace of the transaction execution; 4) Transaction simulation. These RPC methods, such as \texttt{\small eth\_call}, allow users to simulate the execution of a given transaction locally without sending it to the blockchain.

\section{Motivation}
\label{sec:motivation}
This section introduces a real-world client RPC bug and summarizes the limitations of prior studies to motivate our work.

\textbf{Motivating Example.} As introduced in the Section~\ref{sec:intro}, a DoS weakness~\cite{ethcallissue} is reported in the blockchain client Hyperledger Besu. Adversaries can exploit transaction simulation RPC methods such as \texttt{\small eth\_call} and \texttt{\small trace\_call} to make the victim blockchain client execute a resource-consuming smart contract at no cost.

The root cause of this bug lies in a private function located in a deep call chain of the RPC process. As shown in figure~\ref{fig:ethcall}, the Hyperledger Besu client leverages the \texttt{\small calculateSimulationGasCap} function (line 1) to determine the number of gas that a transaction simulation RPC call can consume. When a user provides a \texttt{\small gasLimit} parameter in the RPC method call, the number of gas limit is set to the user-provided gas limit \texttt{\small userProvidedGasLimit} (line 6). This behavior can be triggered if \texttt{\small rpcGasCap} is 0, which is the default value. To exploit this DoS weakness, adversaries can first deploy a gas-consuming smart contract and then call the transaction simulation methods to execute it. Such smart contracts can be super simple, for example, a smart contract containing an infinite loop.

\begin{figure}
    \begin{lstlisting}[language=Java,mathescape, firstnumber=1,escapechar=\%]
private long calculateSimulationGasCap (...) {
    if (userProvidedGasLimit >= 0) {
        if (rpcGasCap > 0) {
            ...
        } else {
            simulationGasCap = userProvidedGasLimit;
}}}
    \end{lstlisting}
    \caption{Java Code Snippet of Hyperledger Besu RPC module}
    \label{fig:ethcall}
\end{figure}

\textbf{Limitations of Prior Work}
Existing works fall short of automatically detecting such RPC bugs due to the following limitations:
\begin{itemize}[leftmargin=*]
    \item Reliance on manual effort. Some existing works aim to detect or exploit RPC bugs using manually crafted rules or test cases. For example, Li et al.~\cite{li2021strong} and Luo et al.~\cite{luo2024towards} proposed exploit schemes for free-contract execution DoS vulnerabilities in blockchain clients. These exploit schemes are highly dependent on expert knowledge and cannot be easily extended to other bug patterns.
    \item Lack of context. Fuzzing methods are also proposed to detect RPC bugs in blockchain clients. EtherDiffer~\cite{kim2023etherdiffer} tests the response consistency among blockchain clients through differential fuzzing. They generated a test blockchain network with a set of predefined smart contracts and transactions targeting them. Therefore, their blockchain context space is limited and cannot cover such gas-consuming smart contracts that trigger the bug.
\end{itemize}

Notably, a set of such context-dependent RPC bugs is also reported in the GitHub repositories of major blockchain clients (Details of these bugs will be provided in \S~\ref{sec:bugdataset}). While some RPC bug contexts can be derived from Ethereum mainnet states, directly identifying these mainnet contexts can lead to scalability issues. Over 79 million smart contracts have been deployed on Ethereum, with over 1 million new transactions deployed every day~\cite{Etherscan}. Therefore, it is important for a detector to support the generation of flexible blockchain state contexts and RPC method calls to detect such RPC bugs automatically.


\section{Design of \ToolName}


In this work, we propose {\small \ToolName}, a context-aware RPC bug detector for Ethereum blockchain clients. We mainly focus on the RPC methods that run locally on a blockchain client, i.e., the blockchain information retrieval, on-chain transaction decoding, and transaction simulation RPC methods as introduced in Section~\ref{sec:background}. We introduce our execution model of the Ethereum client RPC process, as well as the challenges and our corresponding solutions.

\subsection{Execution Model}
\begin{figure}
    \centering
    \includegraphics[width=\linewidth]{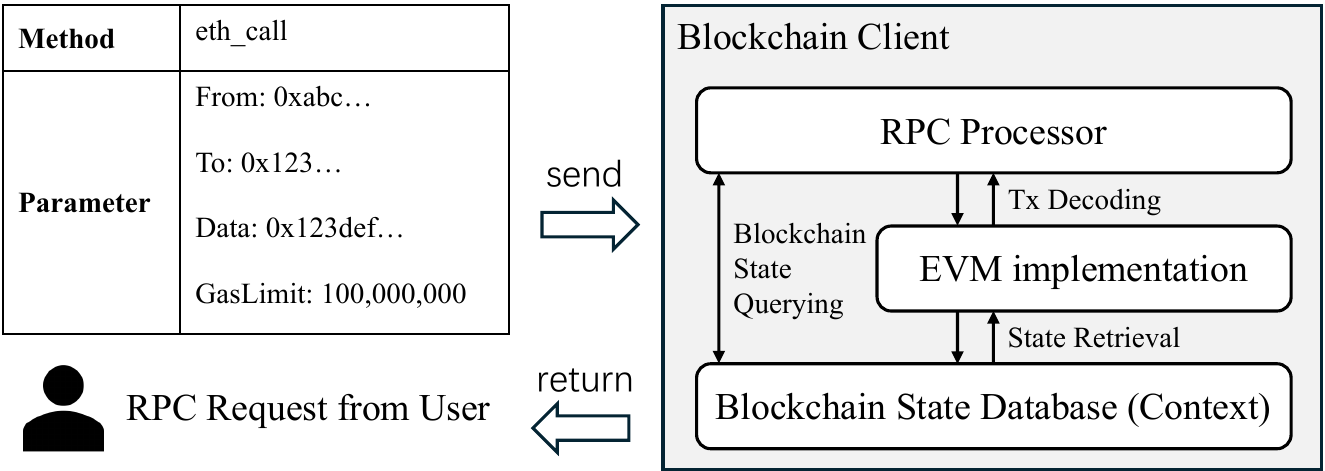}
    \caption{Execution Model of Ethereum Client RPC Process}
    \label{fig:execmodel}
\end{figure}

We propose the Ethereum client RPC execution model as in figure~\ref{fig:execmodel}. As previously discussed, each blockchain client maintains a copy of the blockchain states in its own database. Therefore, the execution of an RPC request under a specific context can be modeled as the execution of the client RPC program with given blockchain states as program variables. If the client receives a blockchain information retrieval request, it will query the storage database and return values to the user. If the client receives a transaction decoding or simulation request, it will invoke its EVM implementation to execute the transaction locally and return the results.

Based on this execution model, the blockchain context space exploration problem can be converted into the client program space exploration problem. Thus, fuzzing techniques based on client code coverage feedback can be leveraged to generate blockchain contexts and invoke RPC method calls.

\subsection{Challenges and Solutions}
\textbf{C1. Blockchain Context Generation.} Ethereum is a transaction-based state machine, where the Ethereum Virtual Machine is leveraged to execute transaction instructions. To fully explore the EVM execution space, automatic tools need to generate smart contracts and transactions that cover as many EVM execution states as possible. However, there are exponential combinations of smart contract opcodes, and most of the opcodes are highly dependent on the stack and memory states generated by previously executed opcodes. The huge combination and dependent relationships make it hard to generate specific smart contracts that can trigger RPC bugs effectively. For example, the hash operation \texttt{\small KECCAK256} takes the stack input as the location of a memory address and calculates the hash of the corresponding data. If there is no pre-executed operation that stores data in the corresponding memory, the operation will just return meaningless values. Besides, if there are not enough stack items, the EVM will encounter the stack underflow exceptions and halt transaction execution. The state-of-the-art transaction generation studies~\cite{yang2021finding, li2024famulet} mainly rely on mutating randomly generated smart contracts. Thus, they can be inefficient in generating valid smart contracts and exploring the transaction execution space.

To solve this challenge, {\ToolName} aims to explore the context generation problem based on on-chain transaction mutation. {\ToolName} first selects a set of transactions from the mainnet that can cover part of the client program space. Since there are millions of new transactions every 24 hours~\cite{Etherscan}, we propose incrementally collecting the transactions of interest that increase the EVM execution code coverage and record them as our initial seed pool. We then perform mutations on these transactions based on EVM execution code coverage feedback. Through these designs, we can improve the efficiency of EVM execution space exploration, thereby enhancing the efficiency of context generation.

\textbf{C2. Method Call Generation.} The main challenge in this part lies in the large parameter space of RPC methods. Blockchain clients provide users with more than 40 RPC methods, each of which accepts various types of input. For example, the transaction simulation RPC method \texttt{\small eth\_call} takes three parameters~\cite{RPCspec}: the transaction to simulate, the simulation block height, and a \texttt{\small stateOverride} object. Each parameter represents a specialized object that consists of various fields of data. The RPC parameters can also contain unstructured inputs. For example, the \texttt{\small eth\_feeHistory} method takes a dynamic array of double values as input to calculate the recorded gas fee of transactions. The unstructured feature of the RPC parameters makes it hard for a detector to generate valid RPC method calls. Besides, the RPC method call that can trigger bugs can be limited in a sub-range of the parameter space. Therefore, it is also challenging for the detector to generate proper RPC method calls that can trigger bugs.

{\ToolName} explores the parameter space of RPC calls by context-aware RPC call generation. We design our RPC call generation based on EtherDiffer~\cite{kim2023etherdiffer}, which defines a Domain Specific Language for each RPC call and its parameter types. {\ToolName} further leverages context information to facilitate parameter generation. For example, the transaction gas limit parameter takes an integer as input, which contains  $2^{32} - 1$ possible values. When mutating this parameter, we slice the integer space into several intervals based on the current block's gas limit. In this way, we can reduce the parameter space to a limited number of intervals.

\textbf{C3. On-chain Execution Overhead.} To create a consistent context for different client implementations, {\ToolName} builds up a test blockchain network consisting of {\NumClient} different client implementations as nodes. However, it also brings extra computational overhead due to the execution of the blockchain network. When deploying new transactions, blockchain nodes need to run auxiliary modules to maintain consensus across the blockchain network. For example, they need to perform signature verification for each transaction. Additionally, there is a delay for the consensus clients in the blockchain network to reach consensus. The transaction mutation process can generate many ``useless'' transactions that do not explore new client program space. Therefore, these auxiliary blockchain modules incur considerable computational overhead and delays when processing unexpected transactions.

To mitigate the on-chain execution overhead, {\ToolName} features a decoupled framework for context generation and method call generation. We design an off-chain transaction simulator to execute transactions without running a blockchain network. Specifically, we leverage the \texttt{\small state\_transition module} of the Go-Ethereum client~\cite{geth} to test the transaction processing module by formatting the context and transaction data as inputs. In this way, we can conduct a specialized test for each of the generated transactions. We only deploy valuable transactions that achieve new code coverage of the client program to the blockchain network. Based on the generated blockchain context, {\ToolName} then conducts RPC calls to the on-chain blockchain clients, ensuring context consistency across different client implementations during testing.

\zjrv{These designs help EthCRAFT detect the bug in the motivating example in several aspects. First, context generation, together with the design to reduce early execution termination, can result in a gas-consuming transaction exceeding the block gas limit. Next, the context-aware RPC call generation can generate a call to simulate this transaction with high gas limit parameters. Eventually, the RPC call is sent to the clients, resulting in inconsistent responses.}

\section{Approach Details}
\label{sec:approach}
\begin{figure}[t]
    \centering
    \includegraphics[width=\linewidth]{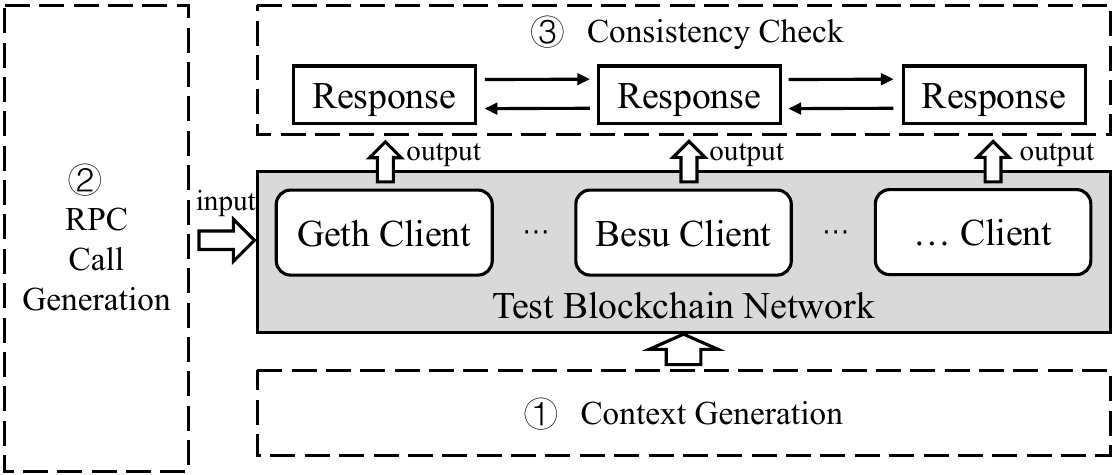}
    \caption{Framework of {\small \ToolName}}
    \label{fig:framework}
\end{figure}

As shown in figure~\ref{fig:framework}, {\small \ToolName} leverages a two-phase fuzzing to detect the RPC bugs in different Ethereum clients. The detection consists of three stages: \circled{\scriptsize 1} Context generation, \circled{\scriptsize 2} RPC method call generation, and \circled{\scriptsize 3} Response inconsistency detection. 

\subsection{Context Generation}

We present the blockchain context generation method used in {\ToolName}. The goal is to generate contexts that can cover as much of the transaction-execution program space as possible. We first propose to select the initial transaction corpus from on-chain transactions. Then we reform these transactions to a semantic-equivalent form to facilitate mutation. Next, we demonstrate our transaction mutation method to explore the transaction execution space. Code coverage feedback is used to guide the mutation .

\begin{algorithm}[t]
	\SetKwInput{KwInput}{Input}
	\SetKwInput{KwOutput}{Output}
 	\SetKwComment{Comment}{//}{}
	\caption{\zjrv{\textbf{Initial Transaction Corpus Selection}}}
	\label{Alg:InitTxSelect}
    \footnotesize
	\KwInput{\footnotesize $\mathbb{T} = \{Tx\}$: Set of Transactions on Blockchain Mainnet; \newline
                \footnotesize $C_{th}$: threshold of accumulated code coverage;
                }
    
	\KwOutput{\footnotesize $\mathbb{R} = \{Tx\}$: Recorded Transaction Corpus Set; \newline
                $\mathbb{O} = \{Opcode\}$: Executed Opcodes Set in $\mathbb{C}$;
                }
        
        $\mathbb{R} \gets \emptyset$ \Comment{\footnotesize Initialize $\mathbb{O}$ to be empty}
        
        $\mathbb{O} \gets \emptyset$ \Comment{\footnotesize Initialize $\mathbb{R}$ to be empty} \label{alg1:empty}
        
        $C_{acc} \gets \emptyset$ \Comment{\footnotesize Initialize accumulated code coverage to be empty}
        
	\While{$C_{acc} < C_{th}$ \textbf{and} $|\mathbb{R}| < 1000$ \label{alg1:noempty}} 
	{
            $Tx \gets \mathbb{T}.\texttt{\footnotesize pop()}$ \label{alg1:getE}\Comment{\footnotesize get a transaction}

            $\mathbb{O}_{tx} \gets \texttt{\footnotesize getOpcodeInSC}(Tx)$ 

            \If{$\mathbb{O}_{tx} \subseteq \mathbb{O}$}
            {
            
            continue \Comment{\footnotesize skip if no new opcode is involved in $Tx$}
            
            }
            
            $C_{tx} \gets \texttt{\footnotesize codeCov}(Tx)$  \Comment{\footnotesize Calculate the code coverage of $Tx$}

            $C_{merge} \gets \texttt{\footnotesize mergeCov}(C_{acc},C_{tx})$
            
            \If{$C_{merge}.\texttt{\footnotesize coverage} > C_{acc}.\texttt{\footnotesize coverage}$}
            {
                $C_{acc} \gets C_{merge}$ \Comment{\footnotesize update accumulated code coverage}
                
                $\mathbb{R}$\texttt{\footnotesize .add}($Tx$)
                
                $\mathbb{O}$\texttt{\footnotesize .add}(executed Opcodes \textbf{in} $Tx$)
            }

	}
	\KwRet{$\mathbb{R}, \mathbb{O}$} \label{alg1:bypass4}

\end{algorithm}

    
    

    
    

        

\subsubsection{Initial Transaction Corpus Selection.}
As there are billions of transactions on blockchain, selecting the transactions of interest is not easy. According to the statistics on etherscan~\cite{Etherscan}, the most popular Ethereum blockchain explorer, there are over 1 million transactions proposed every day during the past two years. Due to resource limits, replaying all the transactions is not feasible. 
Nonetheless, while the transactions can be complex, the execution process can be reduced to iterative patterns. On each execution of the opcodes, the client iteratively checks the runtime stack and memory states and executes the corresponding functions for the opcode. \zjrv{Based on this observation, we propose Algorithm~\ref{Alg:InitTxSelect} to select the initial transaction corpus by incrementally collecting transactions from the blockchain mainnet. The main idea is to maintain a set of triggered OPCODES and randomly inspect transactions if the corresponding smart contract contains new OPCODES.}


\zjrv{Algorithm~\ref{Alg:InitTxSelect} takes a set of on-chain transactions $\mathbb{T}$ during the past two years, and a predefined code coverage threshold $C_{th}$ as input. A set of triggered opcodes is initialized to empty set (line 2). The algorithm starts by randomly selecting a transaction $Tx$ in $\mathbb{T}$ (line 5). We first check whether the involved smart contracts contain new opcodes compared with $\mathbb{O}$ (line 6-7). If not, we skip the transaction and select an alternative (line 9). As different transactions can execute the function call to the same smart contracts, this optimization can help us skip analyzing transactions to contracts that have no new opcodes of interest. There can be cases in which a subcall (i.e., a smart contract calls another smart contract) to a contract in the transaction is determined by the transaction input, which can be missed. Still, we make a trade-off here to improve the efficiency of transaction exploration. }

\zjrv{Next, we use the \texttt{\small debug\_traceTransaction} method to replay the transactions locally and calculate the code coverage of the clients' transaction execution module (line 10). If the transaction covers new code (line 12), we record the transaction in the recorded transaction corpus $\mathbb{R}$, the executed opcodes in the triggered opcode set $\mathbb{O}$, and update the record of covered code of the client (line 13-15). We iterate the above steps and terminate the transaction selection if the cumulative code coverage reaches a preset threshold, or over 1,000 transactions are selected (line 4).}

Although the transaction selection process can incur computation overheads, we note that it is a one-time effort. Once we collect transactions that achieve high cumulative code coverage of the EVM, they can be stored as the initial seed corpus and reused in every round of fuzzing.

\subsubsection{Transaction Reforming.}
The aim of transaction mutation is to explore the transaction execution space based on feedback such as code coverage of EVM execution. Existing works on transaction mutation mainly focus on directly mutating the opcode sequence, such as random insertion and deletion. However, these methods fall short of effectively generating valid smart contracts.

As discussed before, many Ethereum opcodes are highly dependent on the runtime states of stack and memory. A typical type of stack-dependent opcode is the control flow opcodes, i.e., \texttt{\small JUMP} and \texttt{\small JUMPI}. They take the stack input parameter as the location of the jump destination, which is known as the program counter (PC) of the jump destination. A primary constraint on these opcodes is that EVM only allows jumps with a corresponding \texttt{\small JUMPDEST} opcode on the destination. Otherwise, the execution will return invalid. Random mutation can affect the stack input of \texttt{\small JUMP} opcode, as well as the program counter of the \texttt{\small JUMPDEST} opcode. In such cases, the destination of the \texttt{\small JUMP} opcode becomes unpredictable. This can lead to early termination due to the invalid jump destination check, thereby reducing the effectiveness of transaction mutation.

Therefore, ensuring the semantic relationship between the \texttt{\small JUMP} input and PC of the \texttt{\small JUMPDEST} is necessary for generating opcode sequences without being prematurely terminated. 
To mitigate the problem, we propose transaction reforming based on transaction traces, where the transaction traces are ordered sequences of opcodes that were executed in a transaction. 

\begin{figure}
    \centering
    \includegraphics[width=0.7\linewidth]{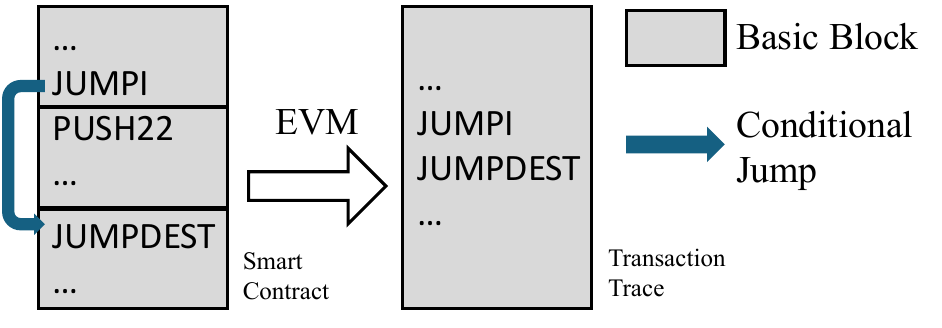}
    \caption{Illustration of the JUMP-JUMPDEST relationship}
    \label{fig:txtrace}
\end{figure}

We first extract the opcode sequence of a target transaction using the aforementioned \texttt{\small debug} method provided by the Ethereum client. As shown in figure~\ref{fig:txtrace}, the trace opcode sequence is a flattened form of the executed part of smart contract opcodes. In the trace opcode sequence, the \texttt{\small JUMP} and \texttt{\small JUMPI} opcodes are not distributed in separate places in different basic blocks of a smart contract. Instead, they always occur consecutively according to the execution trace. Notably, the trace opcode sequence is semantically equivalent to the original smart contract bytecode in the specific transaction, as the EVM executes the same sequence of opcodes during this transaction by definition. Therefore, we can avoid incurring invalid jumps by deleting the adjacent \texttt{\small JUMP} - \texttt{\small JUMPDEST} pairs along with their stack input, while preserving the execution results.

\subsubsection{Transaction Mutation.}
\zjrv{In this step, {\ToolName} mutates the opcode sequence executed in the reformed transaction. The mutation outputs a new opcode code sequence, which will further be deployed as smart contracts on blockchain.} Inspired by existing studies, {\ToolName} leverages two types of mutation strategies based on the mutation granularity towards the opcode sequence: basic block mutation and opcode mutation.

For basic block mutation, we first prepare a set of basic block corpus from the extracted transactions. As discussed in figure~\ref{fig:txtrace}, we eliminated the branches in smart contract opcodes by recording the executed opcodes of given transactions. Therefore, the trace opcode sequence is composed of a continuous set of basic blocks, whose boundary is the \texttt{\small JUMPDEST} opcode. We extract these basic blocks from the selected transactions as corpus. 

During basic block mutation, we employ two mutation strategies: insertion and deletion. Specifically, we start by randomly selecting a location of a basic block in the trace opcode sequence. Then, we choose to delete the corresponding basic block from the sequence or insert a basic block from the corpus. Note that a basic block ending with the \texttt{\small JUMP} or \texttt{\small JUMPI} opcode may cause an invalid jump exception, as discussed before. We delete these opcodes from the basic blocks of the opcode sequence.

We apply similar strategies for opcode mutation. We first maintain an opcodes corpus defined in the EVM specification. During mutation, we randomly select an opcode location in the trace opcode sequence. Then, we employ the opcode insertion and deletion strategies to this location. There are opcodes that take operands as well, such as the \texttt{\small PUSHn} opcode, which will push the operand onto the EVM stack. We also generate a random value for these opcodes.

Our transaction mutation strategy distinguishes itself from existing studies as we combine it with runtime state-aware strategies during mutation. Specifically, we define a number of mutation rules for a special group of opcodes that interact with the runtime states of smart contract storage and memory. For example, the \texttt{\small MLOAD} opcode loads the value stored in a memory location, where the stack input of this operation determines the target memory location. Randomly inserting these opcodes has a high probability of returning the value from a non-initialized memory location. To improve the effectiveness of mutation, we generate valid parameters before inserting these opcodes so that they can produce meaningful outputs. The state-aware mutation is achieved by adding valid storage or memory locations on the stack input before inserting these opcodes. We perform a lightweight static analysis on the opcodes to get the initialized locations in storage and memory.

Specifically, we traverse the opcode sequence with data-flow analysis to record the used storage and memory locations before mutation. We take the following steps: 1) Decompilation of the opcodes. We leverage the Gigahorse decompiler~\cite{grech2019gigahorse,grech2022elipmoc,lagouvardos2020precise} to decompile the opcode sequence to three intermediate representations; 2) Backward data-flow analysis. We start from the last opcode of the sequence and perform a backward traverse. If we encounter an opcode that stores values to storage or memory, i.e., the \texttt{\small SSTORE}, \texttt{\small TSTORE}, and \texttt{\small MSTORE}, we perform a backward dataflow analysis on their parameters based on Gigahorse. We record the locations that can be resolved to values such as operands in a \texttt{\small PUSH} opcode, as well as the program counter of these store operations. The memory location calculation is slightly more complex. By specification, EVM maintains a dynamic pointer for the last memory location that has not been unused, known as the Free Memory Pointer (FMP). Therefore, we maintain the relative memory offset to the FMP for memory operations.

Based on the analysis, we are able to maintain a set of storage or memory locations that are used during the execution. Before inserting opcodes that load these storage or memory locations, we set their parameters to one of the used locations. Although the opcode mutation can change the stack inputs of the state storage opcodes, we notice that the stack input opcode and the storage opcode occur together one after the other in most cases. Therefore, the probability of changing the used storage location is relatively low.

Based on the mutation, we can obtain a new opcode sequence. We then deploy the new opcode sequence as smart contracts on the blockchain, \zjrv{as well as generating a transaction calling this smart contract with the original transaction inputs.} We complete the remaining field of the transaction by randomly filling fields including the value-to-send, the gas price, etc. In this way, Ethereum will execute the opcode sequence contained in the smart contract.
\subsubsection{Code Coverage Feedback.}
We use the code coverage of the client as the metric to evaluate whether a mutated transaction is interesting. To mitigate the computational overhead of the auxiliary blockchain client modules, we design an off-chain transaction simulation tool for the transaction evaluation. We notice that their transaction processing module is designed to follow the specification of Ethereum Yellow Paper. These diverse implementations of transaction processing modules share the same functionalities. Therefore, we can use the code coverage of one Ethereum client as an estimated metric for a transaction to explore the transaction processing program space of other clients.

In this work, we leverage the \texttt{\small state\_transition} module of Go-Ethereum~\cite{geth} (we refer to it as \texttt{\small Geth} for abbreviation), which is designed to process a given transaction and invokes EVM to execute the opcode involved in the transaction. {\ToolName} designs an extension to convert a given transaction into a structured input of this module. Based on this, we call the module in \texttt{\small Geth} and calculate the code coverage within this transaction. If a mutated transaction triggers new code coverages, we then add it to the corpus set for further mutation. In this way, we can avoid invoking other auxiliary modules of the blockchain client, thereby improving test efficiency.

Based on the transaction mutation, {\ToolName} maintains a set of transactions that can achieve considerable code coverage in the client's transaction processing module. We generate a test blockchain network composed of diverse client implementations. The transactions are then sent to the test network to build a context for the test network. 

\subsection{Method Call Generation}

{\ToolName} proposes a context-aware RPC method call generation method to explore the parameter space of RPC methods. The proposed method aims to 1) generate valid parameters for RPC methods so that the RPC request does not be refused by the client; and 2) generate parameters that can trigger different execution conditions during RPC request processing.

To generate valid RPC parameters, we borrow the parameter generation framework proposed by EtherDiffer~\cite{kim2023etherdiffer}. \zjrv{Overall, the DSL contains three main components based on manual annotation for each RPC method: 1) RPC parameter; 2) RPC method call; 3) RPC output format.}

\subsubsection{\zjrv{RPC Parameter}}
\zjrv{This component defines the data types of the parameters and the mutation rules for them. According to the RPC specification~\cite{RPCspec}, the primary data types of RPC parameters consist of \texttt{\small integer} types (e.g., nonce), \texttt{\small address} types (e.g., blockchain address),  \texttt{\small boolean} types, etc. The annotation matches the RPC parameter to these data types based on the specification.  For example, the \texttt{\small eth\_call} method in the motivating example takes a transaction object as input, which can be further divided into primary fields such as the \texttt{\small from} and \texttt{\small to} addresses in \texttt{\small address} type, the \texttt{\small string} type transaction data field, and the \texttt{\small integer} type transferred value.}

\zjrv{{\ToolName} then takes effort to define fine-grained mutation rules for the parameters relevant to the generated contexts. According to the intended input, there are three types of parameters relevant to blockchain contexts: parameter for blockchain state, parameter of blockchain addresses, and parameter of transactions.}

Parameters of blockchain states accept inputs related to certain blockchain states. For these parameters, we first take the corresponding blockchain states as base values, and then we perform mutation based on these base values. For example, the transaction-related RPC calls can take integer parameters such as \texttt{\small maxFeePerGas} and \texttt{\small gasLimit}. For the \texttt{\small maxFeePerGas} parameter, we first record the base fee of the current block (i.e., the minimum gas fee to pay for sending a transaction). Then, we slice the input space of this parameter into a set of intervals by multiplying the base fee with a set of multipliers, such as $\{[0,BaseFee), [BaseFee, 10 \cdot BaseFee), [10 \cdot BaseFee,\infty)\}$. The mutation is performed by randomly selecting an interval and randomly selecting a value in this interval.

The other two types of parameters take in blockchain address and transactions as input. For transaction, we randomly select a recorded transaction that is generated during the transaction mutation process and deployed on the blockchain. For addresses, we randomly select from deployed addresses on the blockchain or generate an empty address as input. There are also parameters that are irrelevant to blockchain contexts. For these parameters, we mainly borrow the random generation method from EtherDiffer.

\subsubsection{\zjrv{RPC method call}}
\zjrv{This component defines how the generated parameters are combined into a valid RPC method call format. Guided by the RPC specifications, the DSL concatenates the names and values of these parameters hierarchically into a \texttt{\small json} format, as well as annotating the RPC method title.}

\subsubsection{\zjrv{RPC output format}}
\zjrv{This component indicates the return data types of the RPC method. Depending on the usage, the return value contains different data structures or errors. For example, the gas estimation RPC method returns the amount of gas used by a transaction. The transaction retrieval RPC method returns a transaction object containing data fields such as the sender, receiver, and transaction data. Therefore, the DSL records the data types of the RPC output format based on the specification. This allows the inconsistency detection module in~\S~\ref{sec:inconsistency} to perform a fine-grained check on each of the data fields in case there are multiple data fields in the RPC return value.}

Based on the above effort, {\ToolName} can generate a set of RPC calls under the current blockchain contexts. \zjrv{Notably, the annotation process for each RPC method does not require much effort, as RPC input and output are very concise and clear according to the specification. Most RPCs require less than 100 lines of code for annotation as a one-time effort. Two of our authors with more than four years of blockchain experience first took a 1-hour training for the annotation process. After that, they spent about 10 minutes per RPC on average.}



\subsection{Response Inconsistency Detection}
\label{sec:inconsistency}
Following the existing studies~\cite{kim2023etherdiffer, yang2021finding, fu2019evmfuzzer}, we leverage the diversity among different blockchain implementations to construct cross-referring bug oracles. Specifically, we send the generated RPC calls to each of the blockchain clients running on our test blockchain network. The responses from these blockchain clients are collected and compared with each other. If any inconsistency is found in their responses, we record the corresponding RPC calls and label them as bugs, as at least one of the clients did not follow the specification.

\section{Evaluation}\label{sec:empirical}

In this section, we evaluate the proposed {\ToolName} and answers the following research questions (RQs):

\begin{enumerate}[leftmargin=*,start=1,topsep=1pt,labelindent=0pt,noitemsep,,label={(\bfseries RQ\arabic*):}]
    \item  How does {\ToolName} perform in detecting the context-dependent RPC bugs?
    \item  How effective is {\ToolName} in detecting new bugs?
    \item  How do the proposed designs improve the performance?
    \item  What are the characteristics of the RPC bugs?
\end{enumerate}

\subsection{Experiment Setup}
To the best of our knowledge, there has been no available dataset regarding the RPC bugs in recent years. The two most relevant client RPC bug datasets are proposed by Yi et al.~\cite{yi2022empirical} and Kim et al.~\cite{kim2023etherdiffer}. In their studies, Yi et al. proposed an empirical study on general client bugs and collected a set of bugs from their GitHub repositories. Kim et al. evaluated their tool on the last version of Ethereum clients at the time and reported a set of RPC inconsistency bugs. However, even the newest bugs involved in these datasets were reported before 2023. The Ethereum network has gone through three main upgrades~\cite{ethhis} after these reports, including the merge hard fork~\cite{merge}, where the consensus mechanism of Ethereum is changed from Proof-of-Work to Proof-of-Stake. These upgrades have greatly changed the functionalities of the blockchain clients, and new bugs can occur after these upgrades.

To evaluate the effectiveness of {\ToolName} regarding the recent client implementations, we build a new benchmark consisting of recently reported RPC bugs regarding {\NumClient} most popular Ethereum client, i.e., Geth~\cite{geth}, Nethermind~\cite{nethermind}, Hyperledger Besu~\cite{besu}, Erigon~\cite{erigon}, and Reth~\cite{reth}. Specifically, we first review the reported issues within these clients' GitHub repositories from January 2023 to January 2025. We filter these issues by the ``bug'' label provided by the repositories to get the reported bug issues. As this work focuses on the RPC bugs of blockchain clients, we use the keyword ``RPC'' to select relevant bug reports from the bug issues. In this way, we have collected 53 issue reports. 

Two of our authors (each of them has over 4 years of blockchain research) then individually read the titles, descriptions, and developer replies in these RPC bug reports to check if they are real bugs. For each of the issues, the two authors will individually label whether it reports a real RPC bug. We then cross-validate their labels to reduce potential bias in our manual classification. If any disagreements arise regarding an issue, we introduce a third author to join the classification and discuss whether the issue reports a real bug. If they cannot reach an agreement, we discard this issue from our benchmark. During the labeling process, we found that some of the issues~\cite{opissue} report RPC bugs out of the Ethereum mainnet, e.g., the Optimism network~\cite{optimism}. Besides, we found that some of the reports are intended behaviors as indicated by the replies from client developers. These reports are out of the scope of this work. Therefore, we filter them out from our benchmark. As a result, we reached a dataset containing 30 RPC bugs. We discarded 23 issues that were either out of our scope or duplicated with existing issues.

We perform two types of evaluation for {\ToolName}: 1) we evaluate {\ToolName} on the reported client versions of these issue reports to check if {\ToolName} can successfully find these bugs. 2) We also run {\ToolName} on the latest version of these Ethereum clients to check if {\ToolName} can find new RPC bugs. All experiments were performed on a Ubuntu 20.04.1 LTS workstation equipped with an Intel i9-10980XE CPU and 256GB RAM.

\begin{table*}
    \begin{tabular}{c l | c c c}
\toprule[0.1em]
        \multirow{2}{*}{\textbf{id}} & \multirow{2}{*}{\shortstack{\textbf{Description of the RPC Bug}}} & \multicolumn{2}{c}{\textbf{Detected ?}}   \\
        & & \textbf{{\ToolName}} & EtherDiffer& \\
        
\hline
 N1 & eth\_call allows unlimited gas usage by default~\cite{ethcallissue} & \checkmark & - & \\
 N2-4 &(Discovered in 3 clients) DoS weakness while retrieving history logs& \checkmark & - & \\
 N5 & \zjrv{eth\_getProof root hash mismatch}~\cite{ErigonTxindex} & \checkmark & - & \\
 N6 & \zjrv{eth\_getTransactionByBlockNumberAndIndex returns inconsistent value}~\cite{NethermindTxindex} &\checkmark & -& \\
 B1 & eth\_getProof return inconsistent value~\cite{B1} & \checkmark & \checkmark &\\
 B2 &eth\_getProof output is off the specification of EIP-1186~\cite{B2} & \checkmark & \checkmark &\\
 B3 & Inconsistency debug\_traceBlock response while tracing genesis block~\cite{B3} & \checkmark & \ding{55} &\\
 B4& eth\_call method does not support EIP-4844 transactions~\cite{B4} & \checkmark & \ding{55} & \\
 B5-6&(Reported in 2 clients) eth\_estimateGas failed to handle EIP-4844 transactions~\cite{B5,B6} & \checkmark & \ding{55} & \\
 B7&Wrong response in trace\_replayBlockTransactions for some failed transactions~\cite{B7} & \checkmark & \ding{55} & \\
 B8&Improper error handling in eth\_getBlockByNumber method~\cite{B8} & \ding{55} & \ding{55} & \\
 B9&eth\_estimateGas ignores some parameters~\cite{B9} & \checkmark & \checkmark & \\
 B10&eth\_call doesn't support integers as block number~\cite{B10}& \ding{55} & \ding{55} &\\
 B11&eth\_call do not accept movePrecompileToAddress parameter~\cite{B11} & \ding{55} & \ding{55} & \\
 B12&eth\_getProof response contain unexpected values~\cite{B12} & \ding{55} & \ding{55} & \\
 B13&Occasionally missing transaction hashes in eth\_getBlockByNumber~\cite{B13} & \ding{55} & \ding{55} & \\
 B14&debug\_getBadBlocks failed to properly handle bad blocks~\cite{B14} & \ding{55} & \ding{55}& \\
 B15-16&(Reported in 2 clients) Inconsistency RPC response in tracing specific transactions~\cite{B15,B16} & \checkmark & \ding{55}& \\
 B17&trace\_block failed when specific transaction in block~\cite{B17}&  \checkmark & \ding{55}& \\
 B18&Inconsistent response in trace\_block Method Output~\cite{B18} & \ding{55} & \ding{55} & \\
 B19&Wrong output for eth\_estimateGas when using create2 opcode~\cite{B19}  & \ding{55} & \ding{55} & \\
 B20&eth\_estimateGas allows unlimited gas limit~\cite{B20}& \checkmark & \ding{55} & \\
 B21&debug\_traceCall fails for specific transactions~\cite{B21} & \checkmark & \ding{55} & \\
 B22&Wrong response from debug\_traceBlockByHash for transactions with BASEFEE opcode~\cite{B22} & \checkmark &  \ding{55} & \\
 B23&eth\_estimateGas does not return decoded revert messages~\cite{B23} & \ding{55}  & \ding{55} & \\
 B24&debug\_traceTransaction return wrong gas cost~\cite{B24} & \ding{55} & \ding{55} & \\
 B25&debug\_traceCall doesn’t return storage keys required~\cite{B25}  & \ding{55} & \ding{55} & \\
 B26&Potential out-of-memory exception when tracing large transactions~\cite{B26} & \ding{55} & \ding{55} & \\
 B27&Potential out-of-memory exception for RPC filter methods~\cite{B27} & \ding{55} & \ding{55} & \\
 B28&Unhandled exception for eth\_getBlockReceipts~\cite{B28} & \ding{55} & \ding{55} & \\
 B29&debug\_traceBlockByHash returns the wrong response when gas is low~\cite{B29}  & \ding{55} & \ding{55}& \\
 B30&eth\_getTransactionReceipt returns the wrong response when reorg happens~\cite{B30} & \ding{55} & \ding{55} & \\
 \hline
 & Total & 20 & 3 & \\
  \bottomrule[0.1em]

    \end{tabular}
    \caption{RPC bug detection results. The bug id started with ``N'' indicates \underline{N}ew bugs found in our study. The bug id started with ``B'' indicates bugs in our \underline{B}enchmark.}
    \label{tab:eval}
\end{table*}

\subsection{Effectiveness of {\ToolName} on the Benchmark}
We first evaluate {\ToolName} on the reported issue benchmark to evaluate the effectiveness of the proposed method. The comparison is made to EtherDiffer~\cite{kim2023etherdiffer}. To the best of our knowledge, EtherDiffer is the only tool that specializes in client RPC bug detection in academic study. For each of the affected client versions, we run {\ToolName} and EtherDiffer for \zjrv{120} minutes and check if they can find the reported RPC bugs in the benchmark.

The detection results are shown in Table~\ref{tab:eval}. From the table, we can find that {\ToolName} outperforms EtherDiffer by successfully detecting more RPC bugs in the benchmark. By inspecting the bugs that {\ToolName} detected, we can observe that {\ToolName} mainly outperforms EtherDiffer in detecting bugs related to transaction decoding and the bugs related to transaction simulation. For example, the RPC bug B7 occurs when a transaction contains a failed call to precompile smart contracts due to insufficient gas provided. EtherDiffer fails to detect it as they do not support the generation of various contexts. This observation also aligns with our motivation to generate various transactions for triggering such context-dependent RPC bugs. Besides, EtherDiffer fails on a set of RPC bugs related to the ``debug'' method, as their tool does not support these RPC methods. Overall, {\ToolName} detected 14 out of the 30 bugs in the benchmark (there are six new bugs not in the benchmark). \zjrv{Notably, we further added a post-hoc analysis of the bug detection time. We found that EtherDiffer detected the three bugs within the first 15 minutes but did not detect the remaining bugs during the remaining time. Similarly, {\ToolName} detected the 14 bugs in the first 30 minutes.}

\zjrv{We further analyze the reasons for the bugs that were not detected by the proposed tool. We found that the RPC bugs, such as B14 and B30, are triggered under contexts outside the execution client. For example, B30 only occurs when there is a chain re-organization in the consensus client, which cannot be covered by EthCRAFT. They also constitute a limitation of the proposed tool. Besides, the RPC bugs, such as B12 and B13, occur very rarely~\cite{B13} and can not be reproduced stably. Some of these bugs even have no clear root causes, and the corresponding GitHub issues remained open for several months~\cite{B13}.}

\subsection{Effectiveness of {\ToolName} on finding new bugs}
We also run {\ToolName} on the latest version of the {\NumClient} clients as of February 2025 to find new bugs. Notably, {\ToolName} has identified six new RPC bugs in major Ethereum clients, as listed in the first four rows of Table~\ref {tab:eval}. These bugs are within the transaction simulation and blockchain information retrieval types of RPC methods. \zjrv{We reported these bugs to the client developers, and all the bugs were confirmed.}

\zjrv{Regarding the severity of the detected bugs in Table~\ref{tab:eval}, 5 of the detected bugs (N1-4 and B20) can be exploited to cause high memory and CPU resource usage of the client node. To trigger N1 and B20, a transaction that consumes more gas than the block gas limit is required as context. To trigger N2-4, specific RPC parameters are required to query a smart contract with multiple storage layouts in the context. In our report to developers, we provided code scripts demonstrating how adversaries could exploit the bugs to trigger node crashes through malicious RPC requests. The corresponding fix of N1 is then written into the \textit{breaking changes} of their GitHub repository's update changelog. The N2-4 bug reports were offered a vulnerability bounty by the Ethereum Foundation.}

\zj{The remaining 15 detected bugs in the table can cause the RPC methods to return incorrect values and cause user confusion. Notably, these RPCs may be used by service providers and DApps, which can affect a wide range of users. For example, bug B15 is reported~\cite{B15} to cause Etherscan~\cite{Etherscan}, the most popular Ethereum blockchain explorer, to display a wrong transferred value in several transactions, undermining the reliability of the service provider.}

\zjrv{We further studied the false positives of the reports. {\ToolName} report inconsistent responses between different client implementations, which might not be bugs. Through investigation, we found there are two types of false positives:}
\begin{itemize}[leftmargin=*]
    \item  \zjrv{The responses are different yet semantically equivalent, which is not restricted by the specification. For example, when calling \texttt{\small eth\_getCode} of a non-existing account, the Geth and Nethermind return “0x”, indicating no code available. Besu returns “null”, which indicates the same conclusion. Notably, the specification does not provide formats for querying the code of a non-existing account. Therefore, we treat this report as a false positive. 16 false positives were found in this category.}
    \item  \zjrv{The developers treat the problem as intended behavior. For example, when calling the \texttt{\small eth\_getBlockByNumber} RPC, Nethermind and Besu return a block object with an additional data field “totalDifficulty”, which is deprecated after the “Merge” update of Ethereum in April 2024. The issue was previously reported~\cite{totaldiff}, but developers of related clients chose not to fix it due to its limited impact. 4 false positives were found in this category.}
\end{itemize}


\subsection{\zjrv{Ablation Study}}
\zjrv{This section evaluates the proposed designs on improving the efficiency and efficacy of blockchain context generation. We compare the performance of EthCRAFT with/without off-chain evaluation, initial transaction corpus selection, and state-aware mutation. For EthCRAFT without (abbreviated as w/o) off-chain evaluation, the generated transaction for each mutation step is deployed to the test blockchain, and we trace its execution in Geth to obtain code coverage feedback. For EthCRAFT w/o initial transaction corpus selection, we use an empty set as the initial corpus. For EthCRAFT w/o state-aware mutation, we skip adding valid storage or memory locations on the stack input before inserting related opcodes.}

\begin{figure}
    \centering
    \includegraphics[width=0.9\linewidth]{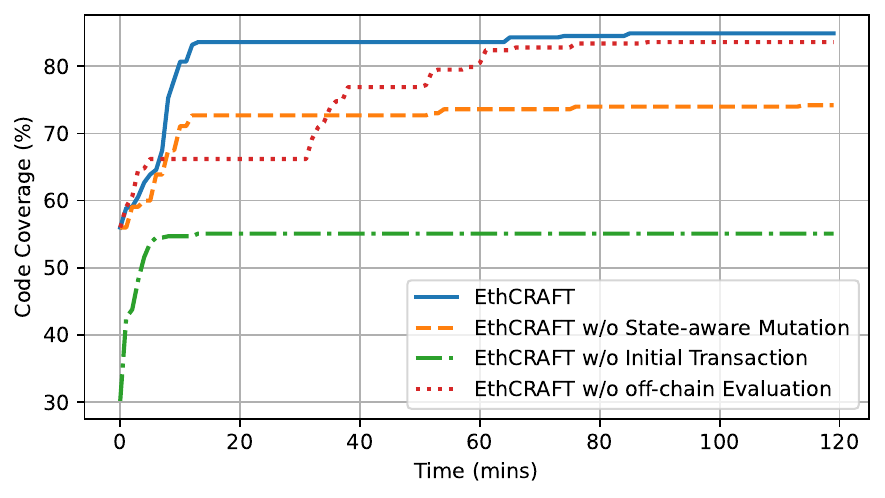}
    \caption{Code Coverage Over Time with/without the Designs.}
    \label{fig:ablation}
\end{figure}

\zjrv{We run the context generation for 120 minutes and use the code coverage of the blockchain client as a metric. The evaluation result is shown in figure~\ref{fig:ablation}. EthCRAFT achieves the highest code coverage and efficiency with all the designs enabled. As a comparison, EthCRAFT w/o Off-chain Evaluation takes more time to achieve high code coverage. The observation aligns with our motivation for off-chain evaluation. On the one hand, the transaction execution result can only be perceived in the next block, which incurs delays in the feedback-based mutation process. On the other hand, on-chain deployment of transactions introduces computational overheads in the blockchain client. Besides, Eth-CRAFT w/o State-aware mutation cannot achieve the highest code coverage. By investigating the mutation results, we found that the inserted memory opcodes often reach invalid states when the current top element on the stack (used to locate a memory location) is very large. The reason is that accessing a large memory location is prevented in Ethereum specification. Additionally, the initial transaction corpus is also crucial for the context mutation to explore the clients' transaction execution program space.}

\subsection{Characteristics of Client RPC Bugs}
\label{sec:bugdataset}
In this research question, we further analyze the characteristics of the detected bugs. We study the distribution and root cause of these bugs. Furthermore, we propose findings from the analysis to facilitate client development.

We first analyze the distribution of RPC bugs among client implementations and the RPC types (introduced in Section~\ref{sec:background}). The distribution of the RPC bugs is shown in table~\ref{tab:dist}. The Info. Ret., Tx Decod., and Tx Simu. in the first row refer to the Information Retrieval, Transaction Decoding, and Transaction Simulation RPC types, respectively. The Cl. refers to Client. The percentage values attached to each client indicate the proportion of each client regarding the number of nodes~\cite{diversity} running on the Ethereum mainnet.

We can get the following observations from the table: 1) The reported RPC bugs are distributed evenly regarding the RPC types. All the RPC types have received 10-15 bug reports. Besides, the reported bugs are also distributed evenly among most clients; 2) A large proportion of bugs are reported in the Hyperledger Besu client. Other clients received a relatively even number of bugs. While the Geth and Nethermind account for more than 35\% of the nodes on the mainnet, the number of bug reports does not exceed much compared with minority clients, i.e., Erigon and Reth.

We find that there are three main root causes of these RPC bugs: 1) Lack of support on new functionalities regarding Ethereum upgrades. For example, RPC bugs such as B4 and B16 in table~\ref{tab:eval} occur because the client's transaction decoding and simulation modules failed to fully follow up on Ethereum Improvement Proposals (EIPs) in the latest Ethereum upgrades. 2) Inconsistency workflow between transaction decoding/simulation and real transaction processing. While these transaction-related RPC methods are expected to execute a transaction locally on the client, they may derive from the real transaction execution due to missing checks on global variables, such as gas. This cause leads to RPC bugs such as B7 and B21. 3) Output format not aligned with the specification. Some of the RPC methods return inconsistent responses with their specification, causing RPC bugs such as B1 and B8. These inconsistencies can confuse users and applications relying on their responses.

          
    

\begin{table}[t]
    \centering
    \resizebox{\linewidth}{!}{
    \begin{tabular}{c | c c c | c}
     \toprule[0.1em]
        \SlashCell{Cl.(\%)}{Type} & Info. Ret. & Tx Decod. & Tx Simu. & Sum \\
        \hline
         Geth (43\%)        &2  &1  &  2     & 5\\ 
         Nethermind (36\%)  &4  &2  & 3       & 9\\ 
         Besu (16\%)        & 6 &4  &   4    & 14\\ 
         Erigon (3\%)       & 3 & 2 &  0      & 5\\
         Reth (2\%)         &0  &2  &  1      & 3\\  
        \hline
        Sum & 15 & 11 & 10 & -\\ 
     \bottomrule[0.1em]
    \end{tabular}}
    \caption{The distribution of the RPC bugs regarding RPC types and client implementations.}
    \label{tab:dist}
\end{table}

\section{Discussion}\label{sec:discuss}

\subsection{\zjrv{Threats to Validity}}
\zjrv{\textbf{Internal Validity.} During context mutation, we use Geth to estimate code coverage, which may miss subtle application logic in other clients. However, since all Ethereum clients follow the same specification for transaction execution to reach consensus, the intermediate state and output of their EVM executions should be the same. Therefore, modeling coverage based on Geth is still a reasonable metric for transaction mutation.}

\zjrv{\textbf{External Validity.} The proposed method aims to detect RPC bugs dependent on on-chain contexts of the Ethereum execution client. We also noticed that several RPC bugs are triggered by context outside the execution client, e.g., chain re-organization due to consensus issues in the consensus client~\cite{reorgBug}, which is outside the scope of our method. We plan to extend our approach to support analyzing consensus contexts as part of our future work.}

\subsection{\zjrv{Limitations}}
\zjrv{\textbf{Extending EthCRAFT to new RPCs and blockchains.} The proposed tool aims to detect bugs for existing RPC methods in Ethereum clients, as Ethereum is one of the largest and most popular blockchain platforms. Extending {\ToolName} to new RPC methods and blockchains requires additional effort, which depends on the type of blockchain. For EVM-based chains, the adaptation requires minor effort. Since they share the same EVM to execute transactions, our context mutation method remains effective. The main effort involves annotating a limited number of new chain-specific RPC methods, which are well-documented. Developers also need to check whether RPC bugs or intended behaviors cause the inconsistency in the report. {\ToolName} cannot be directly extended to blockchains that are not EVM-based, such as Bitcoin~\cite{Bitcoin}, due to their distinct architectures and virtual machines. }


\section{Related Work}\label{sec:related}

\textbf{Blockchain Client Analysis}. Existing works on blockchain client testing and analysis focus on diverse client modules, including EVM execution~\cite{yang2021finding}, consensus protocol~\cite{chen2023tyr,ma2023loki}, transaction handling~\cite{yaish2024speculative,tran2024routing}, rollup layer~\cite{sun2024doubleup,li2024famulet}, etc. Among the studies, Tyr~\cite{chen2023tyr}, Fluffy~\cite{yang2021finding}, and EVMFuzzer~\cite{fu2019evmfuzzer} propose to generate transactions to explore the client program space and peer-to-peer messages to trigger consensus bugs in blockchain networks. These consensus bugs can lead to the violation of the blockchain consensus systems. Wu et al.~\cite{wu2024following} propose to monitor the runtime relationship between the program threads during client execution. LOKI~\cite{ma2023loki} focuses on the bugs in blockchain consensus clients. It performs fuzzing based on a consensus state model of distributed nodes and proposes to generate malicious peer-to-peer messages that undermine the reliability of consensus clients. As for the transaction-handling module, Tang et al.~\cite{wang2024understanding} and Yaish et al.~\cite{yaish2024speculative} studied the attack and defense schemes targeting the transaction pool maintenance mechanism of Ethereum clients. They propose to send low-cost transactions to blockchain clients that can evict benign transactions from the client's transaction pool, causing the Denial of Service attacks.

\textbf{RPC Bug Detection for Blockchain Client}. For client RPC bugs, Yi et al.~\cite{yi2022empirical} empirically studied the patterns of blockchain vulnerability and identified the modules related to these vulnerabilities. They found improper RPC handling can lead to resource-consuming weakness of blockchain clients. DoERS~\cite{li2021strong} studied the free-contract execution vulnerability in blockchain RPC handling and proposed exploitation schemes against RPC service providers. Luo et al.~\cite{luo2024towards} further studied the general Denial of Service weakness in blockchain clients. They proposed a formal verification method to reason the resource model weakness and detect RPC DoS weakness based on the previously proposed free-contract execution vulnerability pattern. These works target a specific type of RPC bugs and rely heavily on the proposed pattern to detect such bugs. Therefore, it is hard for them to extend their detection to other types of RPC bugs. The most relevant work on general RPC bug detection is EtherDiffer~\cite{kim2023etherdiffer}. They start with building a blockchain test network based on a set of pre-defined smart contracts. Furthermore, they propose to randomly generate both semantically-valid and semantically-invalid-yet-executable RPC calls and check the consistency of the responses from different clients. Our work distinguishes from the previous studies as {\ToolName} is able to detect a new and more complex type of RPC bug, i.e., the context-dependent RPC bug. This new type of RPC bug presents unique challenges during context generation and testing, which cannot be easily addressed by simply extending existing methods.

\section{Conclusion}

\zjrv{This paper proposes {\ToolName} for detecting RPC bugs in Ethereum clients. {\ToolName} designs the initial transaction selection method and runtime state-aware mutation strategies to explore the blockchain context space. Based on the generated contexts, {\ToolName} proposes to generate a set of RPC calls with parameters and detect inconsistencies in RPC responses across client implementations. {\ToolName} is evaluated on a set of real-world RPC bugs collected from the GitHub issues of Ethereum clients. The experimental results show that {\ToolName} outperforms existing methods by detecting more bugs. Moreover, our study has reported six new bugs in the major Ethereum client implementation. The experiment's findings indicate that 1) while leveraging the inconsistency between clients’ responses can effectively detect RPC bugs, it also leads to false positives because some inconsistencies may be intended behaviors; 2) RPC bugs can also be triggered by contexts outside the execution client, which can be a possible future research direction for RPC bug detection.}


\begin{acks}
    This work was supported in part by the National Key Research and Development Program of China (No. 2023YFB2704100), the National Natural Science Foundation of China (No. 62572497, No. 624B2139), NSFC-RGC Collaborative Research (No. 62461160332), Guangdong Zhujiang Talent Program (No. 2023QN10X561)
\end{acks}

\bibliographystyle{ACM-Reference-Format}
\bibliography{refs}

\end{document}